\RequirePackage{snapshot}
\documentclass[aps,prb,reprint,superscriptaddress]{revtex4-2}
\usepackage[utf8]{inputenc}

\usepackage[hidelinks]{hyperref}

\makeatletter
\@addtoreset{equation}{myequation}
\@addtoreset{figure}{myfigure}
\@addtoreset{table}{mytable}
\makeatother
\usepackage{hyperref}
\usepackage{color}
\hypersetup{colorlinks=true}
\usepackage{graphicx}
\usepackage{bm}
\usepackage{amsmath}
\usepackage{amssymb}
\usepackage{xspace}
\newlength{\xwidth}

\usepackage[caption=false,labelformat=simple]{subfig}
\usepackage[capitalise]{cleveref}
\usepackage{txfonts}

\usepackage{physics}
\usepackage{siunitx}
\usepackage{chemformula}
\newcommand{\subfigref}[2]{Fig.~\hyperref[#1]{\ref*{#1}#2}}
\usetikzlibrary{external}

\usepackage{wasysym}

\DeclareFontEncoding{LS1}{}{}
\DeclareFontSubstitution{LS1}{stix}{m}{n}
\DeclareSymbolFont{symbols4}{LS1}{stixbb}{m}{it}
\DeclareMathSymbol{\varhexagonblack}{\mathord}{symbols4}{"DD}
\DeclareMathSymbol{\hexagonblack}   {\mathord}{symbols4}{"DE}

\begin{document}
	
	\title{Electrical Non-Hermitian Control of Topological Magnon Spin Transport}
	
	\author{Pieter M. Gunnink}
	\email{pgunnink@uni-mainz.de}
	\affiliation{Institute of Physics, Johannes Gutenberg-University Mainz, Staudingerweg 7, Mainz 55128, Germany}

	\author{Rembert A. Duine}
	\affiliation{Institute for Theoretical Physics and Center for Extreme Matter and Emergent Phenomena, Utrecht University, Leuvenlaan 4, 3584 CE Utrecht, The Netherlands}
	\affiliation{Department of Applied Physics, Eindhoven University of Technology, P.O. Box 513, 5600 MB Eindhoven, The Netherlands}
	\author{Alexander Mook}
	
	\affiliation{Institute of Physics, Johannes Gutenberg-University Mainz, Staudingerweg 7, Mainz 55128, Germany}
	\date{\today}
	\begin{abstract}
		Magnonic topological phases realize chiral edge spin waves that are protected against backscattering, potentially enabling highly efficient spin transport. Here we show that the spin transport through these magnonic chiral edge states can be electrically manipulated by non-Hermitian control. We consider the paradigmatic magnon Haldane model and show that it is transformed into an effective non-Hermitian magnon Chern insulator by including a sublattice-dependent spin-orbit torque. In linear spin-wave theory, this electrically induced torque reduces the damping of the chiral edge magnons along certain edge directions, leading to an enhancement of the spin-wave amplitude. This prediction is confirmed by numerical simulations based on the Landau-Lifshitz-Gilbert equation. 
		For a spin-wave transport setup, in which magnons are excited by a microwave field and detected with a normal metal conductor, we find that the magnon amplification is remarkably robust against disorder, establishing non-Hermitian control as a promising avenue for topological magnonics.
	\end{abstract}
	\maketitle

	\section{Introduction}
	Magnonics is a promising platform for the transport and manipulation of the spin degree of freedom, bypassing the Joule heating associated with conventional electronic devices \cite{chumakMagnonSpintronics2015}. 
	To improve the efficiency of magnon spin transport, it has recently been proposed to exploit magnonic topological phases, such as magnon Chern insulators \cite{katsuraTheoryThermalHall2010,vanhoogdalemMagneticTextureinducedThermal2013,shindouTopologicalChiralMagnonic2013,zhangTopologicalMagnonInsulator2013,mookEdgeStatesTopological2014,owerreFirstTheoreticalRealization2016,kimRealizationHaldaneKaneMeleModel2016,mookInteractionStabilizedTopologicalMagnon2021}, magnon spin Hall insulators \cite{nakataMagnonicTopologicalInsulators2017,mookTakingElectronmagnonDuality2018,kondoMathbb_2Topological2019}, magnon Dirac \cite{franssonMagnonDiracMaterials2016,pershogubaDiracMagnonsHoneycomb2018} and Weyl \cite{liWeylMagnonsBreathing2016,mookTunableMagnonWeyl2016} systems. In particular the magnon Chern insulators, supporting one-dimensional chiral edge modes protected against backscattering, can potentially enable highly efficient spin transport \cite{ruckriegelBulkEdgeSpin2018,wangBosonicBottIndex2020}.

	Magnonic systems naturally couple to their environment. such as through (non-)local dissipation \cite{tserkovnyakSpinPumpingMagnetization2002,tserkovnyakEnhancedGilbertDamping2002,tserkovnyakNonlocalMagnetizationDynamics2005,houshangSpinwavebeamDrivenSynchronization2016,tserkovnyakExceptionalPointsDissipatively2020},	
	non-reciprocal couplings \cite{yuanUnidirectionalMagneticCoupling2023,liReciprocalReservoirInduced2023}, and local pumping through spin-transfer torques \cite{slonczewskiCurrentdrivenExcitationMagnetic1996, bergerEmissionSpinWaves1996}. These couplings are a curse and a blessing at the same time: although they provide additional functionality \cite{gunninkZeroFrequencyChiralMagnonic2023}, they are also responsible for magnon damping, limiting propagation distances.
	
	Herein, we show that the coupling to the environment can also be harnessed to reduce magnon damping. Concretely, we consider a magnon Chern insulator coupled to a metallic layer in such a way that sublattice-dependent spin-orbit torques (SOT) modulate the spin dynamics. We identify a protocol to selectively reduce the damping of the topological chiral edge states.
	
	Furthermore, we study the magnon amplification in disordered systems, showing that the topological protection of the edge states remains, thus allowing for long-distance amplification of spin transport with a much lower effective damping of the edge mode. This is different from previous works which modulate magnon spin transport \cite{anControlPropagatingSpin2014,eveltHighefficiencyControlSpinwave2016,cornelissenSpinCurrentControlledModulationMagnon2018,liuElectricallyInducedStrong2021a}, where the spin transport is not topologically protected and thus sensitive to disorder.
	In addition, we show that our setup realizes a non-Hermitian topological phase, making the connection to the wider field of non-Hermitian topology, where topological features are studied in open systems.  \cite{gongTopologicalPhasesNonHermitian2018,kawabataSymmetryTopologyNonHermitian2019,bergholtzExceptionalTopologyNonHermitian2021,bandresTopologicalInsulatorLaser2018,harariTopologicalInsulatorLaser2018}. The open character of non-Hermitian topological systems not only allows for the fine control of topological features for applications, but also gives rise to topologies not found in Hermitian systems \cite{dingNonHermitianTopologyExceptionalpoint2022}. To realize non-Hermitian topology, magnonic systems offer a promising platform \cite{mcclartyNonHermitianTopologySpontaneous2019a,dengExceptionalPointsSignatures2023,dengNonHermitianSkinEffect2022, flebusNonHermitianTopologyOnedimensional2020, gunninkNonlinearDynamicsNonHermitian2022, hurstNonHermitianPhysicsMagnetic2022,hurstNonHermitianPhysicsMagnetic2022,liMultitudeExceptionalPoints2022,yuNonHermitianTopologicalMagnonics2023,kambojOscillatoryEdgeModes2024}, owing in large part to the ease with which magnons couple to their environment, in combination with the development of magnonic topological phases in the past 15 years \cite{mcclartyTopologicalMagnonsReview2022,zhuoTopologicalPhasesMagnonics2023}. We believe that our findings here demonstrate the versatility of non-Hermitian topology in magnonic systems, opening a pathway towards further on-chip manipulation of spin waves and offering an efficient scheme for the direct electrical control of propagating spin waves.

	This article is organized as follows. In \cref{sec:model} we introduce the non-Hermitian magnon Haldane model and introduce the magnon amplification within linear spin-wave theory. In \cref{sec:numerical} we demonstrate the magnon amplification of chiral edge states using numerical Landau-Lifshitz-Gilbert simulations. In \cref{sec:density} we show that the magnon amplification leads to a signature in the buildup of magnon density. In \cref{sec:transport} we consider a propagating spin wave experiment, and demonstrate that the amplification is robust against disorder. Finally, we end with a conclusion and discussion of the experimental realization in \cref{sec:conclusion}. Additionally, in \cref{app:skin,app:density,app:num,app:transport,app:asym} we show details regarding the hybrid skin-effect, the calculation of the magnon density, the numerical Landau-Lifshitz-Gilbert simulations, the transport calculations and the possibility of an asymmetric spin-orbit torque.

	\section{Model}
	\label{sec:model}
\begin{figure}
		\centering	\includegraphics{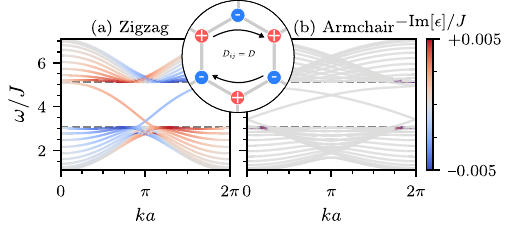}
		\caption{The bandstructure for a zigzag (a) and armchair nanoribbon (b). The colorscale indicates the damping correction $-\Im[\epsilon]$ induced by the spin bias $\mu/J=1$. Note that a positive (negative) damping correction corresponds to an overdamped (amplified) mode. The damping of the left- and right-moving edge mode in the zigzag nanoribbon is enhanced and reduced, respectively. The inset shows the Haldane model used, including a positive and negative SOT on the $\mathcal A$- and $\mathcal B$-sites respectively.
			\label{fig:nh:periodic}}
	\end{figure}

	We consider the magnon Haldane model, a prototypical model of the magnon Chern insulator \cite{owerreFirstTheoreticalRealization2016,kimRealizationHaldaneKaneMeleModel2016}, with the inclusion of a sublattice-dependent SOT, as shown in the inset of \cref{fig:nh:periodic}. The spin dynamics are described by the Landau-Lifshitz-Gilbert equation,
	\begin{equation}
		\partial_t \bm S_i = \bm S_i \times \left(-\frac{\partial \mathcal{H}}{\partial \bm S_i} - \frac{\alpha}{S} \partial_t \bm S_i + \frac{\alpha_{\mathrm{sp}}}{S}\bm S_i \times  \bm\mu_i\right),
		\label{eq:nh:LLG}
	\end{equation}
	where $\alpha \equiv \alpha_0 + \alpha_{\mathrm{sp}}$ is the sum of the Gilbert damping $\alpha_0$ and the interfacial Gilbert damping enhancement $\alpha_{\mathrm{sp}}$ \cite{tserkovnyakEnhancedGilbertDamping2002}. Throughout we set $\hbar=1$. The Hamiltonian is given by
	\begin{equation}
		\mathcal H = -\frac{1}{2} \sum_{ij} \left[J_{ij} \bm S_i \cdot \bm S_j - D_{ij} \hat{\bm z} \cdot\left(\bm  S_i \times \bm S_j\right) \right] 
		-  H_0\sum_i  S_i^z,
		\label{eq:nh:ham}
	\end{equation}
	where nearest neighbors experience an exchange coupling, $J_{ij}=J$, and next-nearest neighbors are coupled through the Dzyaloshinskii-Moriya interaction (DMI), $D_{ij}=-D_{ji}=D$. The spins are aligned to an external magnetic field applied in the $z$ direction, contributing a Zeeman energy $H_0$. Furthermore, $\bm\mu_{i}=\mu_i\, \hat{\bm{z}}$ is the spin accumulation in the normal metal attached to site $i$, taken such that
	\begin{equation}
		\mu_{i}=\begin{cases}
			+\mu & i \in \mathcal{A} \\
			-\mu & i \in \mathcal{B}
		\end{cases}
		\label{eq:nh:stt}
	\end{equation}
	changes sign between sublattices $\mathcal{A}$ and $\mathcal B$. We refer to $\mu$ as the spin bias throughout this work. 
	For the magnon Haldane model in Eq.~\eqref{eq:nh:ham} with its ground state spin texture oriented out of the plane, the anomalous spin Hall effect  \cite{dasSpinInjectionDetection2017,dasEfficientInjectionDetection2018} can be used to create a spin accumulation at the normal metal to ferromagnet interface. Other realizations of a magnon Chern insulator, such as in-plane field-polarized Kitaev-Heisenberg magnets \cite{zhangTopologicalMagnonsThermal2021}, may require the spin Hall effect to induce the spin bias.

	We linearize the LLG equation~\eqref{eq:nh:LLG} in deviations $m_i=(S_i^x + \mathrm{i} S_i^y)/\sqrt{2S}$ from the uniform state $\bm S_i=S\hat{\bm z}$, apply the Fourier transform of the spin-wave operators, $m_{\mathcal{A/B},i} = \sqrt{2/N} \sum_{\bm k} \mathrm{e}^{\mathrm{i} \bm k\cdot \bm R_i}m_{\mathcal{A/B},\bm k}$ and obtain the equation of motion,
	\begin{equation}
		\mathrm{i}(1+\mathrm{i}\alpha)\partial_t \bm \Psi_{\bm k} = \bm{\mathcal{H}}_{\bm k}\bm{\Psi}_{\bm k},
		\label{eq:nh:eom}
	\end{equation}
	where we have introduced the effective non-Hermitian Hamilton matrix
	\begin{equation}
		\bm{\mathcal{H}}_{\bm k} =   (H+3JS)\sigma_0 + \bm h_{\bm k} \cdot \bm\sigma + \mathrm{i} \gamma\sigma_z.\label{eq:nh:hamK}
	\end{equation}
	Here $\bm \Psi_{\bm k} = (m_{\mathcal{A},\bm k}, m_{\mathcal{B},\bm k} )^T$ is the magnon state vector, $\bm \sigma$ is a vector of Pauli matrices and 
	\begin{equation}
		\bm{h}_{\bm k} = S \sum_i 
		\begin{pmatrix}
			-J \cos(\bm{k} \cdot \bm{\delta}_i) \\
			J \sin(\bm{k} \cdot \bm{\delta}_i)\\
			2D \sin(\bm{k} \cdot \bm{\rho}_i)
		\end{pmatrix},
	\end{equation}
	where $\bm \delta_i$ and $\bm \rho_i$ are the vectors connecting nearest and next-nearest neighbors respectively. 
	Furthermore, we have incorporated the SOT in the Hamiltonian, resulting in an effective imaginary mass $\mathrm{i} \gamma\sigma_z$, with $\gamma\equiv\alpha_{\mathrm{sp}}\mu$, which renders the Hamiltonian non-Hermitian. The damping of a magnon with frequency $\omega$ will then be given by the sum of (i) the Gilbert damping, $\alpha\omega$ and (ii) a \emph{damping correction} due to the spin-orbit torque, $-\Im[\epsilon]$, where $\epsilon$ is the complex energy obtained from diagonalizing the effective non-Hermitian Hamiltonian $\bm{\mathcal{H}}_{\bm k}$.

	The stability of this system can be determined by requiring that  $\Im[\omega_{\bm k}]<0$  for all $\bm k$. Thus, expanding $\omega_{\bm k}$ around $\bm k=0$, we obtain 
	\begin{equation}
		\hbar\omega_{\bm k =0} = H + 3JS - i\alpha H - \sqrt{9J^2S^2-\gamma^2},
	\end{equation}
	we obtain that the system is stable if $\gamma^2<9J^2S^2$. At $\gamma^2=9J^2S^2$ there is an exceptional point, signaling an instability \cite{dengExceptionalPointsSignatures2023}.
	Additionally, at the Dirac points, $\bm K=(0,4\pi/3\sqrt{3}a)$ and $\bm K'=(2\pi/3a,2\pi/3\sqrt{3}a)$, $\omega^\pm_{\bm k}$ is given by
	\begin{align}
		\hbar\omega^\pm_{\bm K}&=\left(H+3JS\pm3\sqrt{3}DS\right)(1-i\alpha) \pm i\gamma \\ \quad \hbar\omega^\pm_{\bm K'}&=\left(H+3JS\pm3\sqrt{3}DS\right)(1-i\alpha) \mp i\gamma,
	\end{align}
	where $\pm$ refers to the upper and lower magnon band. We thus obtain the additional stability requirement 
	\begin{equation}
		\gamma/\alpha < H+3JS-3\sqrt{3}|D|S.
	\end{equation}
	

	The Chern number $\mathcal C_n$ of the $n$-th band is still well-defined in the presence of the imaginary mass $\gamma$, and we find $\mathcal C_{1}=-1$ and $\mathcal{C}_2=1$, if $D>0$ and $|\gamma|/JS<1$ \cite{liGainLossInducedHybridSkinTopological2022}. From the bulk-boundary correspondence it thus follows that in the topologically non-trivial phase there exist chiral edge modes for open boundary conditions. However, the finite imaginary mass $\mathrm i\gamma$ will lead to crucial modifications of the chiral edge modes' damping and localization, resulting in reduced damping and localization of the modes on one side of the sample \cite{zhuHybridSkintopologicalModes2022}.  
	
	In what follows, we set $S=1$, $H=1.1$, $\alpha_0=5\times10^{-3}$ and $\alpha_{\mathrm{sp}}=5\times10^{-3}$. 
	Furthermore, we set $D/J=0.2$ and $a=\SI{3.5}{\angstrom}$, inspired by the candidate magnonic topological material \ch{CrI3} \cite{chenTopologicalSpinExcitations2018}.
	
	We show in \cref{fig:nh:periodic} the dispersion of a nanoribbon, chosen with either zigzag or armchair edges along the periodic boundary conditions. The colorscale indicates the \emph{damping correction}, where a positive (negative) damping correction corresponds to an overdamped (amplified) mode.
	Note that the total damping of each mode is positive, i.e., $-\Im[\epsilon]<\alpha\omega$ for all modes, as required for stability.
	We emphasize that our calculations include the fact that boundary spins have a lower coordination number than the bulk spins, effective reducing the on-site potential on the edge \cite{pantaleonEffectsEdgeOnSite2018,pershogubaDiracMagnonsHoneycomb2018a}.

	For the zigzag edges, \subfigref{fig:nh:periodic}{(a)}, the right-moving modes have reduced damping, i.e., $-\Im[\epsilon]<0$, because they have support on the $\mathcal A$-sites, with a positive spin bias applied. In contrast, the left-moving modes have an increased damping, because of their support on the $\mathcal B$-sites, with a negative spin bias applied. Right-moving edge modes are therefore amplified relative to left-moving edge modes. 
	The bulk modes also have damping corrections, but since these are not topologically protected, disorder will cut down their lifetime, while the edge modes remain protected, as we will show below in the context of spin transport below.

	For the armchair-terminated nanoribbon, \subfigref{fig:nh:periodic}{(b)}, edge modes are not amplified and the tiny damping corrections to the bulk modes are too small to be visible on the chosen colorscale. Instead, for the armchair edges, the effect of the SOT manifests itself as a hybrid skin-effect, localizing the edge modes on one side of the sample \cite{kunstBiorthogonalBulkBoundaryCorrespondence2018,okumaTopologicalOriginNonHermitian2020,zhuHybridSkintopologicalModes2022,liGainLossInducedHybridSkinTopological2022,borgniaNonHermitianBoundaryModes2020a}, as further analyzed in \cref{app:skin}.

	\section{Numerical LLG simulations} 
	\label{sec:numerical}
\begin{figure}
		\includegraphics[width=\columnwidth]{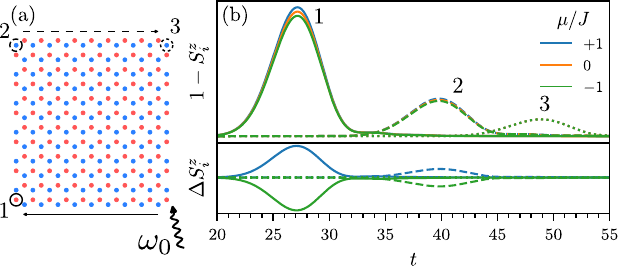}
		\caption{(a) The rectangular system considered, where a spin wave is excited with a pulse at frequency $\omega_0$ at the bottom right corner. (b) The spin-wave amplitude, $1-S_i^z$, is shown, for sites ``$1$'' (solid), ``$2$'' (dashed) and ``$3$'' (dotted) indicated in (a). The difference $\Delta S_z\equiv S_i^z-S_i^z|_{\mu=0}$ is shown below, i.e, the difference as the spin bias is turned on. For $\mu>0$, the zigzag edge indicated by a solid/dashed arrow amplifies/suppresses. The armchair edges do not amplify nor suppress.
			\label{fig:nh:square} }
	\end{figure}

	Having established the magnon amplification within linear spin-wave theory, we next confirm it by means of numerical simulations of the full LLG in finite-size systems, as indicated in \subfigref{fig:nh:square}{(a)}. We initialize the system in the uniform state $\bm S_i=S\hat{\bm{z}}$ and excite an edge mode at time $t=0$ on the bottom right corner, with a frequency $\omega_0/J=3.8$ in the gap. We track the spin-wave amplitude, defined as $1-S^z_i$, and the spin-wave amplitude difference under bias, $\Delta S_z\equiv S_i^z-S_i^z|_{\mu=0}$, at selected sites at the edge at a distance from the excitation point. Below, we refer to these measuring points, denoted by ``1, 2, 3'' in \subfigref{fig:nh:square}{(a)}, as ``detectors.''
	Further details of the simulations are discussed in \cref{app:num}.

	The simulated system is a rectangle, with approximately similar lengths of armchair and zigzag edges and we show the detected signal in \subfigref{fig:nh:square}{(b)}. The excited edge mode travels around the system, arriving at detector ``1'' after a characteristic time determined by its group velocity. As the edge mode has traveled through a zigzag edge, its signal is either amplified or suppressed, depending on the sign of the spin bias $\mu$ [compare green and blue lines in \subfigref{fig:nh:square}{(b)}], leading to a signature in $\Delta S_i^z$. Next, going from detector ``1'' to ``2'', the mode passes through an armchair edge, which does not amplify the mode. Thus, while the signal at detector ``2'' is reduced by Gilbert damping, $\Delta S_i^z$ stays nonzero. Finally, the modes arrives at detector ``3'' after traveling through a zigzag edge with an opposite termination to the first zigzag edge. Therefore, the modes that were previously amplified are now suppressed, and vice versa, as quantified by a zero $\Delta S_i^z$ in \subfigref{fig:nh:square}{(b)}. 
	
\begin{figure}
		\centering
		\includegraphics[width=\columnwidth]{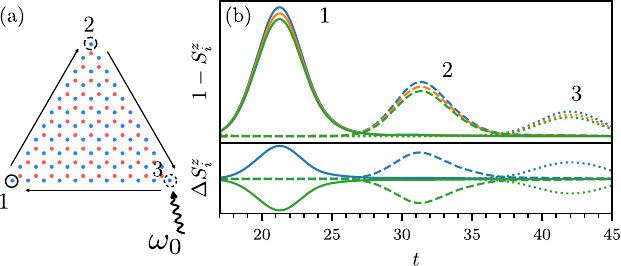}
		\caption{
			The same as \cref{fig:nh:square}, but for a triangle configuration chosen such that all edges are of the zigzag type \emph{and} amplifying for $\mu>0$.
		}
		\label{fig:triangle}
	\end{figure}

	It is also possible to design a triangle oriented such that all of its edges are of the zigzag type {and} terminated predominantly by $\mathcal{A}$-sites. Therefore, each edge amplifies the mode, resulting in a recurrent amplification as the mode travels around the system. This setup is shown in \cref{fig:triangle}, where the splitting is present at all detectors. Note however that due to the finite Gilbert damping in the system, the growth is bounded. 

	\section{Magnon density} 
	\label{sec:density}
	The identified magnon amplification leads to a signature in the average magnon density $\langle n_i\rangle$. To show this, we add a stochastic magnetic field $\bm h_i$ to the LLG equation~\eqref{eq:nh:LLG} that enforces the quantum-mechanical thermal population of magnons \cite{brataasHeatTransportAntiferromagnetic2015,benderEnhancedSpinConductance2017,zhengGreenFunctionFormalism2017}. We set $\mu/J=\pm0.002$, such that the spin accumulation in the attached normal metal is smaller than the lowest magnon band ($|\mu|\ll H$), and also small compared to the temperature, which we expect to apply to any real system at room temperature. For the case where $|\mu|\approx H$, see \cref{app:density}.

\begin{figure}	\includegraphics[width=\columnwidth]{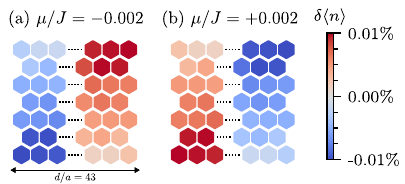}
		\caption{The relative change in magnon occupation, $\delta \langle n \rangle$, for opposite values of the spin bias $\mu/J$: (a) $-0.002$ and (b) $+0.002$, summed over a hexagon containing three $\mathcal A$ and three $\mathcal B$ sites. The sign of spin bias determines the localization of the edge modes on the top or bottom zigzag edge. The total length of the nanoribbon is $d/a=43$, and we only show the outermost hexagons. \label{fig:nh:heatmap} }
	\end{figure}

	In Fig.~\ref{fig:nh:heatmap} we show the relative change in the magnon occupation, $\delta\langle n\rangle$, between the cases with and without spin bias, for a nanoribbon of length $d/a=43$. Depending on the sign of the spin bias $\mu$, the magnon density increases ($\delta\langle n\rangle>0$, red) or decreases ($\delta\langle n\rangle<0$, blue) on opposite corners of the lattice [compare Figs.~\ref{fig:nh:heatmap}(a) and (b)], which is a direct result of the amplification of the zigzag edges that transport spin towards these corners. As shown in \cref{app:density}, the modulus of $\delta \langle n\rangle $ is smaller for systems with shorter zigzag edges, and saturates around $d^*/a =50$, since for zigzag edges much shorter than the magnon relaxation length, a single magnon spends as much time on the top as on the bottom zigzag edge, thereby not experiencing a net amplification effect. This interpretation is further supported by the observation that $d^*$ decreases for increasing Gilbert damping (i.e., for a decreasing relaxation length), as further discussed in \cref{app:density}.
	
	We conclude that the magnon density is a good measure of the amplification, especially since the spin accumulation can be electrically controlled---allowing a single experimental setup, potentially using nitrogen-vacancy magnetometry \cite{rondinMagnetometryNitrogenvacancyDefects2014}, to switch between the cases presented in Fig.~\hyperref[fig:nh:heatmap]{\ref*{fig:nh:heatmap}(a)} and \hyperref[fig:nh:heatmap]{(b)}.

	\section{Transport} 
	\label{sec:transport}
\begin{figure}
		\centering
		\includegraphics[width=\columnwidth]{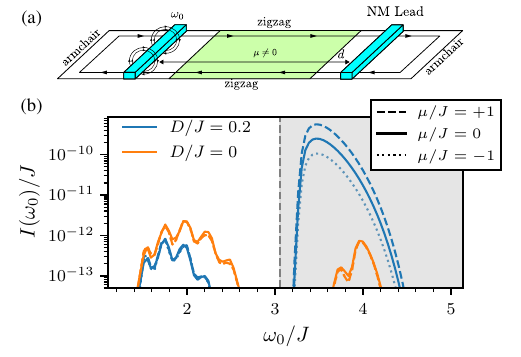}
		\caption{(a) The transport setup considered. Magnons are excited with a transversely oscillating magnetic field in the left antenna and are detected through spin pumping into a normal metal lead on the right. We consider the transport through zigzag edges, which exhibit amplification. (b) The injected spin current in the right lead, $I(\omega_0)$, as a function of excitation frequency $\omega_0$, at a fixed distance $d/a=200$ comparing the topologically trivial ($D=0$) and topologically non-trivial ($D/J=0.2$) case and the effects of $\mu$. The shaded area indicates  the topological magnon gap supporting the chiral edge states. The bulk transport is suppressed by the presence of disorder in these calculations.}
		\label{fig:magnon-transport}
	\end{figure}

	Since the chiral edge magnons are robust against elastic back scattering, they are particularly interesting for highly efficient spin transport \cite{ruckriegelBulkEdgeSpin2018}. Below, we show that the topological edge spin transport can be electrically controlled in \emph{disordered} systems with zigzag termination. We consider a propagating spin wave experiment as shown in \subfigref{fig:magnon-transport}{(a)}. Magnons are excited with a microwave antenna---modelled by a local excitation field for sites $i$ below the antenna, $\bm h_i^{\mathrm{exc}}=h\,(\cos(\omega_0 t),\sin(\omega_0 t), 0)$ with strength $h$ and frequency $\omega_0$---and detected with a normal metal strip, into which they inject spin. The spin bias is non-zero only between the injector and the detector [see green region in \subfigref{fig:magnon-transport}{(a)}], such that detection and excitation are performed in the Hermitian part of the system. We include disorder as random on-site magnetic fields drawn from a uniform distribution in the interval $[-\delta/2,\delta/2]$.
	We consider a nanoribbon of length $d$, zero temperature, and a finite disorder level of $\delta/J=1$ (such that, since $H=1.1$, all magnon energies remain positive). 
	For further technical details of the transport calculation, and results for varying levels of disorder, see \cref{app:transport}.

	\begin{figure}
		\centering
		\includegraphics[width=\columnwidth]{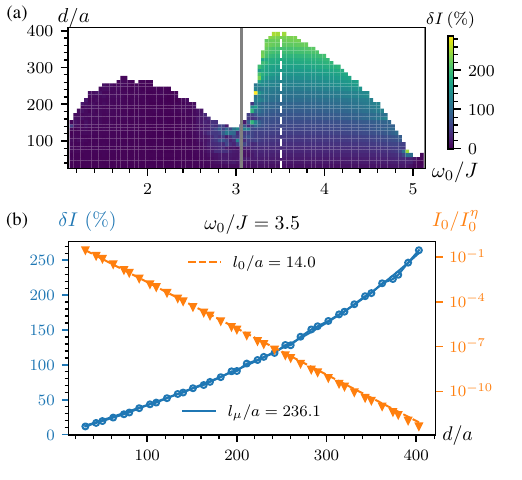}
		\caption{ (a) For $D/J=0.2$ and $|\mu|/J=1$, the relative amplification $\delta I\equiv (I^{+\mu}-I^{-\mu}) / 2I^0$, as a function of distance, $d$, and excitation frequency, $\omega_0$. We only show $\delta I$ if $I^{+\mu}>10^{-12}I_{\eta}$, where $I_{\eta}$ is the maximum transported spin current at $d/a=30$ and $\mu=0$. (b) At frequency $\omega_0/J=3.5$ [indicated by the white dashed line in (a)] the relative amplification $\delta I$ as a function of $d$ (left axis, open symbols). The solid lines are a fit to the function $\sinh(d/l')$. Also shown is the spin current $I_0$ relative to the maximum transported spin current $I_0^\eta$ at $d/a=30$, for no spin bias (right axis, solid triangles), fitted to $\exp(-d/l_0)$ (solid line).}
		\label{fig:amplification-factor}
	\end{figure}
	
	The resulting spin current injected in the right lead is shown in \subfigref{fig:magnon-transport}{(b)} for $\mu/J=\pm1$ and $\mu=0$, for the topological trivial ($D=0$) and non-trivial ($D/J=0.2$) system. Firstly, we observe that in the topologically trivial case, $D=0$, there is no notable amplification of the signal for $\mu/J=\pm1$ compared to $\mu=0$. 
	However, for the non-trivial case, $D\neq0$, we observe an amplification or damping, depending on the sign of the spin bias $\mu$. This result is in direct agreement with the amplification found in the numerical LLG simulations (recall \cref{fig:nh:square}). 
	The amplification or damping is strongest for excitation frequencies in the gap, where the topologically protected edge modes reside. We conclude that the amplification of the edge modes is robust against disorder, while the amplification or damping of the bulk modes is strongly suppressed, and can be effectively ignored.

	To investigate the distance dependence of the amplified signal we define the relative amplification factor $\delta I\equiv (I^{+\mu}-I^{-\mu}) / 2I^0$ as the amplification relative to the unmodified signal. \subfigref{fig:amplification-factor}{(a)} shows $\delta I$ as a function of $d$ and $\omega_0$. For numerical stability, we only plot $\delta I$ if $I^{+\mu}>10^{-12}I_{\eta}$, where $I_{\eta}$ is the maximum transported spin current at $d/a=30$ and $\mu=0$.
	Only the edge modes exhibit sizable relative amplification, as expected from their disorder immunity; their $\delta I$ increases with distance, since the longer an edge magnon travels, the more $\mathcal A$-sites with positive SOT it travels over. The bulk modes on the other hand have scattered off impurities before any amplification can take hold.

	As shown in \subfigref{fig:amplification-factor}{(b)} (orange line), the transported spin current follows an exponential decay, $I^{\pm\mu}\propto\exp(-d/l_0)\exp(\pm d/l_\mu)$, where $l_0$ is the spin-bias-independent decay length, and $l_\mu$ is the amplification length scale, with $l_\mu>l_0$ (see \cref{app:transport}). Consequently, we obtain $\delta I=\sinh{(d/l_\mu)}$, which fits the numerical data very well (blue line). From this fit, we have determined $l_\mu/a=231.8$ and $l_0/a=14$ at $\omega_0/J=3.5$. These numbers agree with the approximations $l_0\approx v_g/(2 \alpha \omega_0) \approx 14a\approx \SI{5}{nm}$ and $l_\mu\approx v_g/(\alpha_\text{sp}\mu) \approx 200a \approx\SI{70}{nm}$, where $v_g = Ja$ is the group velocity of the edge mode.
	Importantly, $l_\mu$ is inversely proportional to $\mu$, and thus for a weaker SOT the amplification grows slower with distance. For example, if $\mu/J=0.1$, we obtain $l_\mu\approx10^{3}a\approx\SI{500}{\nano m}$. 
	From \subfigref{fig:amplification-factor}{(b)} one can now read off the expected relative amplification factor $\delta I$ for a distance $d$ where the unamplified signal can still be measured, e.g., for $d/a=300$ we find $\delta I\approx160\%$ and a decay of $I_0/I_0^\eta\approx 10^{-9}$.

	\section{Conclusion and experimental realization} 
	\label{sec:conclusion}
	We have shown that the chiral edge states in the magnon Haldane model can be electrically controlled through applying a SOT. For the zigzag edge geometry this results in an amplification of the edge modes, which we have confirmed using numerical Landau-Lifshitz-Gilbert simulations. In addition, this amplification is reflected in the magnon density, which gets increased on one side of the sample. Finally, we have shown this enhanced transport to be robust against disorder within the linear spin-wave theory formalism, indicating that amplification over large distances is a possibility. Throughout, we have assumed a sublattice-antisymmetric SOT [recall Eq.~\eqref{eq:nh:stt}], but we show in the \cref{app:asym} that the amplification of the edge modes is qualitatively the same for a sublattice-\textit{asymmetric} SOT, which we believe to be experimentally easier to realize. 
	
	Based on our results, we foresee two possibilities to realize the non-Hermitian topological magnon phase considered in this work: (i) certain magnetic compounds and (ii) artificial magnetic materials. First, the sublattice-dependent SOT can be engineered by putting a spacer between the magnetic and normal metal layer that breaks the sublattice symmetry or by using magnetic layers with a built-in sublattice asymmetry, e.g., due to buckling as realized in honeycomb or kagome materials \cite{marshallEu2Mg3Bi4CompetingMagnetic2022, gibsonMagneticElectronicThermal2023}. A normal metal layer placed above and below the magnetic layer would then couple asymmetrically to the $\mathcal A$ and $\mathcal B$ sites respectively. Applying a positive spin accumulation $\mu$ in the top normal metal and a negative spin accumulation $\mu$ in the bottom normal metal, for example by opposite voltages, then realizes the antisymmetric SOT as considered in this work. 
	Secondly, the sublattice-dependent SOT could also be realized in artificial magnetic materials, such as topological magnonic crystals \cite{shindouChiralSpinwaveEdge2013,shindouTopologicalChiralMagnonic2013} and magnetic solitons in a honeycomb lattice \cite{kimChiralEdgeMode2017}. These artificial materials would offer remarkable control over both the driving and the edge geometry.  
	
	Finally, we comment here on the size of the spin-orbit torque. With sufficient optimization, the spin-orbit torque $\gamma$ can be \SI{10}{mT} per \SI{e11}{A m^{-2}} \cite{fukamiMagnetizationSwitchingSpin2016,zhangMagnetotransportMeasurementsCurrent2013}.
	The maximal current density that can be applied is device specific, but as an example we take here magnonic waveguides composed of Bi-doped Yttrium Iron Garnet (BiYIG) and Platinum (Pt), where current densities of \SI{e11}{A m^{-2}} have been achieved in the context of magnon amplification \cite{merboucheTrueAmplificationSpin2024}. Similar current densities have been achieved in other systems \cite{eveltHighefficiencyControlSpinwave2016}. 
	Taking a a current density of \SI{e11}{A m^{-2}}, we obtain a spin-orbit torque of $\gamma=\SI{10}{mT}$. For $\alpha_{\mathrm{sp}}=\num{5e-3}$ we then obtain $\mu=\SI{0.2}{meV}$. Given that typically $J\approx \SI{1}{meV}$, this gives a ratio of $\mu/J=\num{0.2}$. Following the analysis as presented in \cref{sec:transport}, this corresponds to $l_\mu=\num{e3}a\approx\SI{350}{nm}$.
	
	\begin{acknowledgments}
		This work is in part funded by the Fluid Spintronics research program with Project No.~182.069, financed by the Dutch Research Council (NWO), and by the Deutsche Forschungsgemeinschaft (DFG, German Research Foundation) -- Project No.~504261060 (Emmy Noether Programme).
	\end{acknowledgments}
	
	%

	
	
	%
	
	\appendix

	\section{Hybrid skin-effect}
	\begin{figure}
		\centering
		\includegraphics[width=\columnwidth]{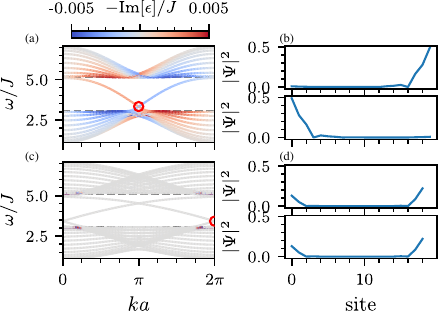}
		\caption{For a zigzag (top row) and armchair nanoribbon (bottom row), we show the magnon dispersion (a,c), the localization of the eigenmodes circled in red (b,d). The colorscale in (a,d) indicates the damping enhancement, $-\Im[\epsilon]$ induced by the spin bias $\mu/J=1$. The damping of the left- and right-moving edge mode in the zigzag nanoribbon is enhanced and reduced respectively, whereas in the armchair nanoribbon the edge modes exhibit the hybrid skin-effect and thus have larger support on one side of the sample. 
			\label{fig:nh:periodic-full}}
	\end{figure}
	\label{app:skin}
	Besides a damping correction, the model considered also exhibits the hybrid skin-effect, which we discuss here in more detail. Importantly, this hybrid skin-effect manifests itself only for nanoribbons with an armchair edge. For nanoribbons with a zigzag edge, the edge modes have symmetrical support on either the top or bottom: the left-moving mode is localized on one side, the right-moving mode on the other side of the ribbon. To demonstrate this, we show in \subfigref{fig:nh:periodic-full}{(a, b)} the zigzag-terminated magnon spectrum and the localization of the eigenmodes. We obtain a left-moving mode and right-moving mode, which are localized on opposite sides of the sample. 
	
	For the armchair ribbon, \subfigref{fig:nh:periodic-full}{(c, d)}, the eigenmodes $|\psi|^2$ have an asymmetric distribution in space, and have stronger support on one side of the sample. Therefore, the armchair edge modes exhibit a \emph{hybrid} skin-effect. The skin effect generally refers to the localization of eigenmodes on one side of the sample \cite{kunstBiorthogonalBulkBoundaryCorrespondence2018,okumaTopologicalOriginNonHermitian2020}, but here the localization is only present for edge modes in the armchair geometry. We therefore refer to this as the hybrid skin-effect, as suggested by Refs.~\cite{liGainLossInducedHybridSkinTopological2022,zhuHybridSkintopologicalModes2022}, who first discovered this effect. Not shown here are the bulk modes, but we have confirmed that these do not exhibit this asymmetric localization. Finally, we note that the asymmetric localization of the magnon density discussed in the main text, is the equivalent of the hybrid skin-effect discussed here, but for open boundary conditions \cite{liGainLossInducedHybridSkinTopological2022}.
	
	In the zigzag orientation there is no hybrid skin-effect, but there is a damping enhancement of the edge modes---which has the same physical origin, namely the non-Hermitian topology. Both effects are therefore two sides of the same coin, manifested differently under different boundary conditions.

	\section{Details of calculations for the magnon density}
	\label{app:density}
	\begin{figure*}[t]
		\centering
		\includegraphics[width=\textwidth]{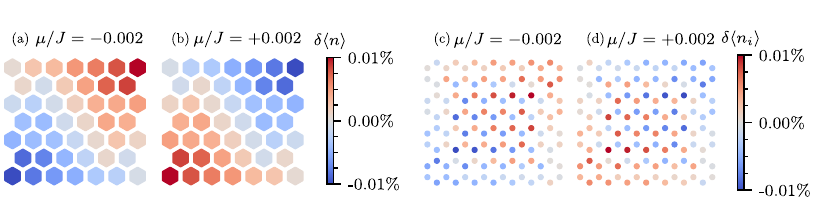}
		\caption{(a-b) The relative change in magnon occupation, for a smaller finite nanoribbon, with length $d/a=7$, such that we are able to show the entire nanoribbon. (c-d) The relative site-resolved change of magnon density, before spatial averaging. The magnon density changes between the $\mathcal A$ and $\mathcal B$ sites, and thus the build up of spin is hard to observe. After averaging one sees the build up of spin more clearly, as in (a-b). Parameters for (a-d) are identical to Fig.~\ref{fig:nh:heatmap} in the main text. }
		\label{fig:hexagons-and-averaging}
	\end{figure*}
	
	To calculate the magnon density, we add to the LLG equation a magnetic field $\bm h_i$, modelling stochastic fluctuations, such that $\partial_t\bm S_i|_{\mathrm{stoch}}=-\bm S_i \times \bm h_i$. After linearization, we obtain the equation of motion 
	\begin{equation}
		\mathrm i(1+\mathrm i\alpha)\partial_t m_i = \sum_j H_{ij} m_j, \label{eq:eom}
	\end{equation}
	where
	\begin{equation}
		H_{ij} = \delta_{ij}\left(H+S\sum_n J_{in}+\mathrm{i}\gamma_j\right) - S(J_{ij} + \mathrm{i} D_{ij})\label{eq:Hij}
	\end{equation}
	is the linear spin-wave Hamiltonian and $\gamma_i\equiv\alpha\mu_i$. 
	We now Fourier transform the equation of motion to frequency space, to obtain
	\begin{equation}
		\sum_j \mathbb G _{ij} ^{-1} (\omega) m_j(\omega) = h_i^{0}(\omega)+h_i^{\mathrm{sp}}(\omega). \label{eq:Geom}
	\end{equation}
	Here, $h_i^{0/\mathrm{sp}}(\omega)$ is the Fourier transform of the circular components $h_i^{0/\mathrm{sp}} = h_i^x + \mathrm{i} h_i^y$ of the stochastic magnetic field, taking into account the fluctuations related to the bulk Gilbert damping ($h_i^{0}$) and the interfacial spin-pumping ($h_i^{\mathrm{sp}}$). The inverse magnon propagator is given by $\mathbb G _{ij} ^{-1} (\omega) = -\delta_{ij}(1+\mathrm{i}\alpha)\omega + H_{ij}$,
	where $H_{ij}$ is the effective non-Hermitian Hamiltonian in real space, given by Eq.~\eqref{eq:Hij}. 
	At finite temperatures, the stochastic magnetic field $h_i(\omega)$ has to be chosen such that $\expval*{h_i(\omega)}=0$ and $\expval*{h_i^{0/\mathrm{sp}}(\omega)h^{0/\mathrm{sp}}_j(\omega')^*}=2\pi\delta(\omega-\omega')\mathbb R_{ij}^{0/\mathrm{sp}}(\omega)$, where 
	\begin{equation}
		\mathbb R_{ij}^0(\omega) = \delta_{ij} \frac{4\alpha_0\omega / S }{e^{\omega/k_BT}-1}\quad\mathrm{and}\quad \mathbb R_{ij}^{\mathrm{sp}}(\omega)  = \delta_{ij} \frac{4\alpha_{\mathrm{sp}}(\omega-\mu_i) / S }{e^{(\omega-\mu_i)/k_BT}-1} \label{eq:Rij}
	\end{equation}  
	are covariance matrices determined by the quantum-mechanical fluctuation-dissipation theorem to ensure agreement with the quantum-mechanical linear spin-wave theory for magnons \cite{brataasHeatTransportAntiferromagnetic2015,benderEnhancedSpinConductance2017,zhengGreenFunctionFormalism2017}.
	
	\begin{figure*}
		\centering
		\tikzsetnextfilename{magnon_density_length}
		\includegraphics{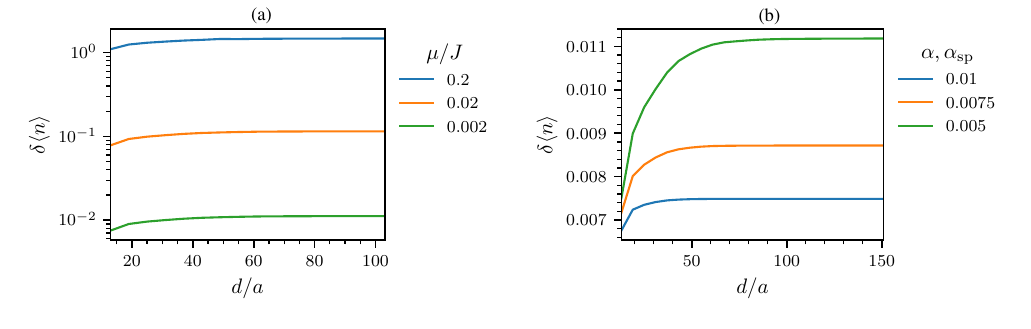}
		\caption{(a) The change in magnon density for the upper-right hexagon, as a function of length $d$ of the nanoribbon, for different values of the spin bias $\mu$. (b) Same as (a), but for different values of $\alpha=\alpha_{\mathrm{sp}}$, while keeping the product $\gamma=\alpha_{\mathrm{sp}}\mu$ constant at $\gamma/J=10^{-5}$. Note that the decrease in magnitude for increasing $\alpha$ can be explained by the corresponding decrease in $\mu$ to keep $\alpha_{\mathrm{sp}}\mu$ constant.}
		\label{fig:magnon-density-length}
	\end{figure*}

	The average magnon number, $\expval{n_i}$, of the local spin $i$, can then be found as
	\begin{equation}
		\expval{n_i} =\int\frac{\dd{\omega}}{2\pi} \left\{\mathbb G(\omega)\left[\mathbb{R}^0(\omega)+\mathbb{R}^{\mathrm{sp}}(\omega)\right]\mathbb G^\dagger(\omega)\right\}_{ii}.
	\end{equation}
	To highlight the relative change of $\langle n_i\rangle$ under the spin bias $\mu$, we show in the main text \begin{equation}
		\delta \langle n_i\rangle\equiv \frac{ \langle n_i\rangle_{\mu=0}-\langle n_i\rangle}{\langle n_i\rangle_{\mu=0}}\times100\%,    
	\end{equation}
	i.e., the relative change of magnon density as the spin bias is turned on. Because the site-resolved occupation is strongly dependent on the on-site spin bias, we average over a hexagon containing three $\mathcal A$ and three $\mathcal B$ sites.  We note that one single site can be part of up to three hexagons.    To show that the averaging procedue does not neglect information, we show in \subfigref{fig:hexagons-and-averaging}{(a-b)} the change in magnon density for a smaller nanoribbon with $d/a=7$, after and before the averaging procedure.

	We consider a nanoribbon of length $d$ and width of $6a$, oriented such that its zigzag edges are of length $d$. In the main text, we only show the change in magnon density for the three outermost hexagons. 
	We find that the change in magnon density increases as a function of nanoribbon length. We show this in \subfigref{fig:magnon-density-length}{(a)}, where we find that for small ($d/a<50$), the change in magnon density increases, saturating for larger sizes. The sizes for which saturation is reached is found to be independent of the spin-orbit torque strength. We have therefore chosen to show in the main text the case for $d/a=43$, for which the change in magnon density is already $95\%$ of the change in magnon density at $d/a=103$. Additionally, we show in \subfigref{fig:magnon-density-length}{(b)} the change in magnon density for increasing Gilbert damping, where we have set $\alpha=\alpha_{\mathrm{sp}}$, whilst keeping $\gamma=\alpha_{\mathrm{sp}}\mu$ constant. For increasing Gibert damping, the saturation length reduces. Additionally, the overall magnitude of the change in magnon density increases, but this we attribute to the fact that we keep $\gamma$ constant---such that as we increase $\alpha_{\mathrm{sp}}$, we simultaneously decrease $\mu$ to keep the product $\alpha_{\mathrm{sp}}\mu$ constant.

	
	\begin{figure}
		\centering
		\includegraphics[width=\columnwidth]{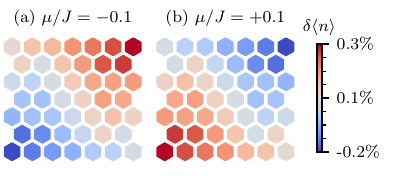}
		\caption{The relative magnon density with $\mu/J=0.1$ and $T/J=1$, for opposite values of the spin bias $\mu/J$. Other parameters are identical to Fig.~\ref{fig:nh:heatmap} in the main text. Importantly, the overall average magnon density increases and is no longer symmetric around $\delta\langle n\rangle=0$, which we attribute to quantum-mechanical corrections to the FDT, since here $\mu$ is comparable to the temperature $T$.}
		\label{fig:heatmap-large-mu}
	\end{figure}

	The stochastic magnetic fields in Eq.~\eqref{eq:Rij} follow from with the quantum-mechanical FDT. We expect the quantum-mechanical nature of the FDT to become relevant if the spin accumulation is comparable to the temperature energy scale. In the main text, we have chosen $|\mu|/J=0.002$, and we therefore do not expect such corrections there. To demonstrate that the quantum-mechanical nature of the FDT becomes relevant for large $\mu$, we choose $\mu/J=0.1$ (as a reminder, $T/J=1$), and show the resulting magnon density [for a small nanoribbon] in Fig.~\ref{fig:heatmap-large-mu}. We observe that the magnon density on average increases for both positive and negative spin bias $\mu$, which we therefore attribute to the quantum-mechanical FDT. In realistic systems at room temperature however, we do not expect the spin accumulation to be comparable to the temperature energy scale, and we thus concentrate on $\mu/T\ll1$ when discussing the magnon density in the main text.

	\section{Details of numerical Landau-Lifshitz-Gilbert simulations}
	\label{app:num}
	We describe here the details for numerically solving the Landau-Lifshitz-Gilbert equation [Eq.~\eqref{eq:nh:LLG} in the main text]. A spin wave is excited with a pulse of a local transversely oscillating magnetic field $\bm h_{e} =b_0 f_e(t)(\cos\omega_0t\, \hat{\bm{x}}+\sin\omega_0t\, \hat{\bm{y}})$ added to the LLG equation, such that $\partial_t \bm S_i|_{e}=-\delta_{in}\bm S_i \times \bm h_e$, where $n$ is the excitation site. Here $f_e(t)=\exp[-(t-b)^2/2c^2]$ is an envelope function which slowly turns the pulse on and off, since turning on the excitation pulse instantaneously will excite a range of spurious frequencies. We have chosen $b=4.5 J^{-1}$, $c=3J^{-1}$ and $b_0/S=5\times10^{-3}$. 
	At $t=0$ all spins are aligned along the $z$ directions, and the LLG equation is numerically solved with timesteps of $\Delta t = 0.01 J^{-1}$. 
	
	\section{Details of transport calculations}
	\label{app:transport}
	
	We consider in the main text a transport setup as shown in \subfigref{fig:magnon-transport}{(a)}. Magnons are excited by an antenna on the left side, travel through a nanoribbon oriented such that the transport occurs parallel to the zigzag edge, and inject spin into an attached normal metal right side. We model this by taking a nanoribbon of length $d$, and add a local driving field $h_i^{\mathrm{exc}}=b_0\cos\omega_0 t\, \hat{\bm x}\,\delta_{i\in \mathrm{antenna}}$ to the leftmost sites. To the rightmost sites we add a normal metal lead. In order to minimize reflections, we add an interfacial Gilbert damping enhancement $\alpha_\mathrm{IF}$ to the left- and rightmost sites. For the sites between the left antenna and right normal metal lead we allow for $\mu\neq0$, whilst $\mu=0$ for the sites in contact with the left antenna and right normal metal. The detection and injection therefore happens in the Hermitian phase, but transport happens in the non-Hermitian phase. We consider $T=0$ in these calculations, but note that at finite temperatures there will also be spin injected in the lead because of the thermal population of the magnons. However, this effect can easily be subtracted experimentally.

	Similar to Eq.~\eqref{eq:Geom}, the equation of motion for this system is given by 
	\begin{equation}
		\sum_j \mathbb G _{ij} ^{-1} (\omega) m_j(\omega) = h_i^{\mathrm{exc}}(\omega),
	\end{equation}
	where $h_i^{\mathrm{exc}}(\omega)=h\delta(\omega-\omega_0)\delta_{i\in \mathrm{antenna}}$ is the Fourier transform of the excitation field.
	
	The spin current injected in the right lead can then be found from the continuity equation $\partial_t \langle S_i^z\rangle=0$ as \cite{ruckriegelBulkEdgeSpin2018} 
	\begin{equation}
		I(\omega_0) = \alpha_\text{IF}\int \frac{\dd{\omega}}{2\pi} \sum_{i\in \mathrm{right\ lead}} 
		\left\lbrace\omega\mathbb{G}(\omega) \mathbb H (\omega) \mathbb{G}^\dagger (\omega)\right\rbrace _{ii},
	\end{equation}
	where $[\mathbb H (\omega)]_{ij}= h_i^{\mathrm{exc}} (\omega) h_j^{\mathrm{exc}} (\omega)=h^2\delta(\omega-\omega_0)\delta_{i\in \mathrm{antenna}}$, the summation is over all lattice sites in contact with the right lead and $\alpha_\text{IF}$ is the interfacial Gilbert damping enhancement of the normal metal lead serving as the detector. In order to minimize reflections, we choose the Gilbert damping enhancement of the antenna and detection lead as $\alpha_\text{IF}=1$. 
	
	In addition to the uniformly distributed on-site disorder $\delta$ discussed in the main text, here we also investigate the effect of defect disorder. We accounted for defect disorder by randomly removing a fraction $w$ of the sites, implemented by setting a large on-site magnetic field $H_{\mathrm{defect}}=1000H$ on those sites. In addition, the spin-orbit torque is set to zero on those sites, to ensure that they are truly defects. We average over $N=100$ realizations.
	
	\subsection{Effects of disorder}
	\label{sec:relative-delta-I}
	
	\begin{figure*}%
		\centering
		\tikzsetnextfilename{disorder-relative}
		\includegraphics{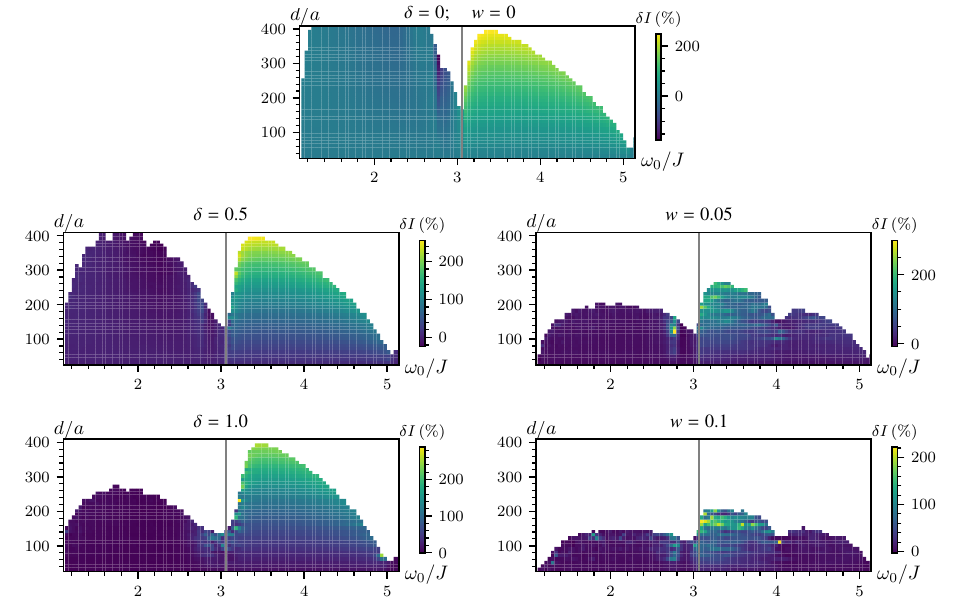}
		\caption{The relative spin transport amplification factor $\delta I$ as a function of antenna distance $d$ and excitation frequency $\omega_0$ for selected disorder concentrations. $\delta$ denotes the random on-site potential, and $w$ the defect concentration. (Top) No disorder, $\delta = w = 0$. (Left panels) Only random on-site potential disorder. (Right panels) Only defect concentration. For numerical stability, we only plot $\delta I$ if $I^{+\mu}>\nu I_{\eta}$, where $I_{\eta}$ is the maximum transported spin current at $d/a=30$ and $\mu=0$, and $\nu=10^{-12}$ for the random on-site potential disorder and $10^{-10}$ for the defect concentration.
			In both cases, for increasing disorder, the bulk signal is suppressed, while the signal carried by the edge modes is more robust. Note that the color scales are not equivalent for each figure, and that the color scale extends to negative values. However, $\delta I$ is always positive for excitation of the edge modes.}
		\label{fig:disorder-relative}
	\end{figure*}

	We first study the effect of disorder on the observed relative amplification factor $\delta I\equiv (I^{+\mu}-I^{-\mu}) / (I^{+\mu}+I^{-\mu})$, shown in Fig.~\ref{fig:disorder-relative}. We first observe that in the absence of disorder, both the bulk and edge modes amplify the signal, with both positive and negative sign of $\delta I$. Importantly, the edge mode amplification factor is always positive, whereas the bulk mode amplification factor can have either sign, depending on excitation frequency and distance $d$ to the antenna. Upon adding disorder, the bulk amplification factor is significantly reduced, whilst the edge signal is relatively unaffected. Increasing the disorder $\delta$ significantly suppresses the relative amplification factor $\delta I$ in the bulk, whilst the edge mode is relatively unaffected. This we attribute to the topological protection of the edge modes, which disallows backscattering. Additionally, we observe some smaller differences between the on-site disorder $\delta$ and defect concentration $w$. Most strikingly, the defect concentration suppresses the relative amplification factor for larger $\omega_0$, whereas the on-site disorder does not. Furthermore, the defect concentration does not produce as clear a signal as the on-site disorder, which we attribute to the fact that for the same number of realizations, the defect concentration has not yet converged. 
	
	\subsection{Model of relative amplification factor $\delta I$}
	
	\begin{figure*}
		\centering
		\includegraphics{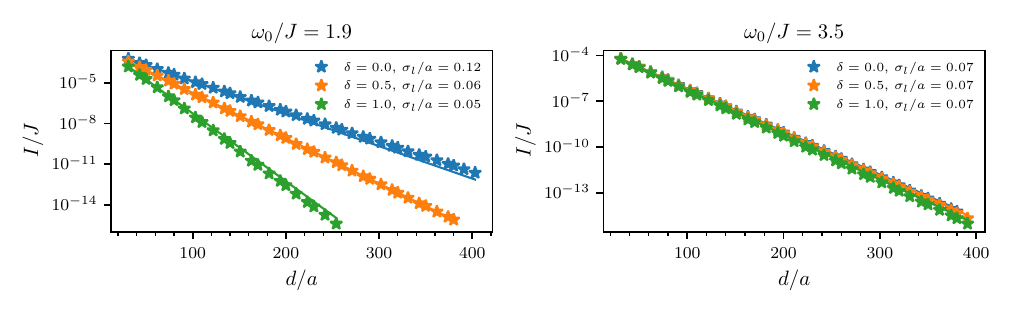}
		\caption{The injected spin current with $\mu/J=+1$, as a function of distance, for a frequency in the bulk (left) and a frequency in the band gap (right). Note that the injected spin current is shown on a $\log$-scale.  The signals are represented by the stars, and the lines are numerical fits to \cref{eq:exp-decay}, where $\sigma_l$ is the standard error of the $l$ parameter, defined as the square root of the covariance.  }
		\label{fig:signals_linecut}
	\end{figure*}
	
	\begin{figure*}
		\centering
		\includegraphics{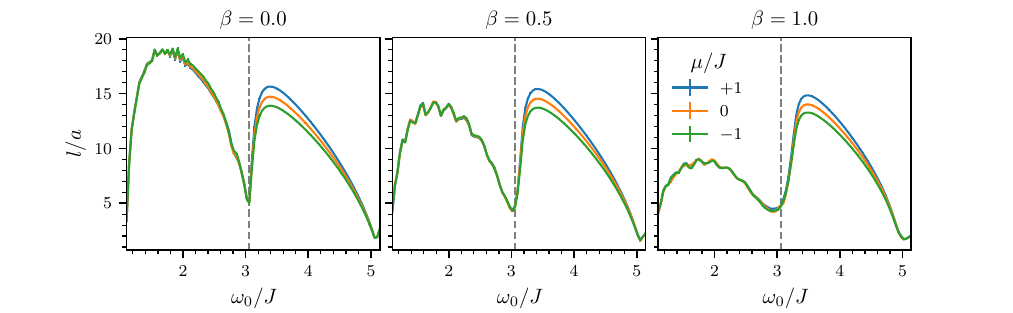}
		\caption{For increasing disorder $\beta$, the fitted decay length $l$, as defined in \cref{eq:exp-decay}. The dashed line indicates the lower band gap, separating bulk and edge modes. Error bars indicate standard error of the $l$ parameter, defined as the square root of the covariance, but most errorbars are too small to be seen. }
		\label{fig:fit_data}
	\end{figure*}

	To further explain the distance dependence of the relative amplification factor, we propose here a simple model, where we assume that the injected spin currents decay exponentially \cite{ruckriegelBulkEdgeSpin2018},
	\begin{equation}
		I^{\mu}(\omega_0) = c_0 \exp\left( -\frac{d}{ l}\right), \label{eq:exp-decay}
	\end{equation}
	where $l^\mu$ is the effective decay length of the excited magnon with frequency $\omega_0$ and applied spin orbit torque $\mu$, and $c_0$ is a constant. We will first show that this model of exponential decay is accurate, by performing a numerical fit, as shown in \cref{fig:signals_linecut}. We choose a frequency in the bulk and in the band gap, such that we excite a bulk and edge mode, and perform a numerical fit to \cref{eq:exp-decay}. We stress here that the data shown here is after averaging over $N=100$ realizations of the disorders. In addition, we show in \cref{fig:fit_data} the result of performing this fitting procedure to all frequencies and disorder levels. In both figures, we show the standard error of the fitted $l$, defined as the square root of the covariance. From the quality of these fits we can conclude that the assumption of exponential decay is well justified, even in the presence of an applied spin bias $\mu$.
	
\begin{figure*}
		\centering
		\tikzsetnextfilename{lprime}
	\includegraphics{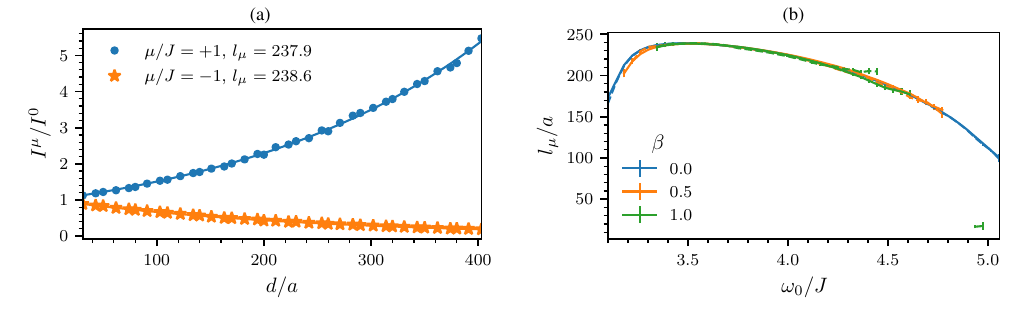}
		\caption{(a) For $\omega_0/J=1.9$, the ratio of injected spin current with a spin bias $\mu\neq0$ to no spin bias $\mu=0$. The solid lines indicate the resulting fit to $\exp\left( \mp{d}/{l_\mu}\right)$. (b) For frequencies in the gap, the fitted $l_\mu$ for positive (solid lines) and negative (dashed lines) spin bias $\mu$. The errorbars show the standard error, and we only show $l_\mu$ if the standard error is less than $5a$. Importantly, we conclude that $l_{+\mu}=-l_{-\mu}$. }
		\label{fig:lprime}
	\end{figure*}

	We now proceed to further develop a model for the effective decay length, which we propose can be written as 
	\begin{equation}
		l = v_g \tau,
	\end{equation}
	with $v_g$ the group velocity of the mode with frequency $\omega_0$ and the lifetime
	\begin{equation}
		\tau^{-1} = \tau_0^{-1} + \tau^{-1}_\mu,
	\end{equation}
	where $\tau_0$ is the spin bias-independent lifetime, due to a combination of Gilbert damping and scattering of defects, and $\tau_\mu$ is the lifetime enhancement as a result of the applied spin bias $\mu$. 
	We can thus write the injected spin current as 
	\begin{equation}
		I^{\mu}(\omega_0) \propto \exp\left( -\frac{d}{ l_0}\right) \exp\left( -\frac{d}{ l_\mu}\right), \label{eq:exp-decay-lprime}
	\end{equation}
	where $l_0\equiv v_G \tau_G$ and $l=v_G\tau_\mu$ is the amplification length scale. In \subfigref{fig:lprime}{(a)} we show the ratio of $I^{\mu}/I^0$ for a single frequency and disorder level. Following \cref{eq:exp-decay-lprime}, we fit this data with the assumption $I^{\pm\mu}/I^0\propto \exp\left( \mp{d}/{l_\mu}\right)$, and show the resulting fit as the solid lines. In \subfigref{fig:lprime}{(b)} we perform this fit for all frequencies in the band gap and show the resulting $l_\mu$. The errorbar indicates the standard error of the fitted $l$, and we only show datapoints where the standard error is smaller than $5a$. We can now draw two conclusions: (i) the amplification or suppressing is well described with the model of exponential decay with two lifetimes and (ii) for opposite spin bias, the effective amplification length scale $l_{\pm\mu}$ has opposite sign, and can thus be well approximated as $l_{+\mu}=-l_{-\mu}$.

	We are now in a position to develop a model for the amplification factor, $\delta I\equiv (I^{+\mu}-I^{-\mu}) / 2I^0$, which can be written as
	\begin{equation}
		\delta I = \frac12 \left( e^{-\frac{d}{l_{+\mu}}} - e^{-\frac{d}{l_{-\mu}}} \right) = \sinh\left({\frac{d}{l_\mu}}\right).
	\end{equation}
	where we have assumed that $l_{+\mu}=-l_{-\mu}$.
	
	We can additionally perform a simple estimate, by taking the edge magnons in the middle of the gap, for a ferromagnet with compensated boundaries \cite{owerreFirstTheoreticalRealization2016}, such that we have $v(\omega_0)= Ja$, and by assuming $\tau_\mu = 1/ ( \alpha_\text{sp}\mu)$. For the parameters $\mu/J=1$ and $\alpha_{\mathrm{sp}}=\num{5e-3}$, as in the main text, we then find $l_\mu/a=200$. Additionally, for the edge modes, the lifetime $\tau_0$ is independent of disorder, as can be concluded from \subfigref{fig:lprime}{(b)}. Thus, we have $\tau_0=(2\alpha\omega_0)^{-1}$, and obtain $l_0=v_g\tau_0\approx 14$ for $\omega_0/J=3.5$. Both these estimates are in excellent agreement with the $l_0$ and $l_\mu$ obtained from fitting the numerical results, as shown in \cref{fig:amplification-factor} in the main text.

	\section{Asymmetric spin-orbit torque}
	\label{app:asym}
	To contrast the scenario of sublattice-\emph{antisymmetric} spin bias studied in the main text, we here consider the case of sublattice-\emph{asymmetric} spin bias. Concretely, we assume that the spin accumulation is non-zero on the $\mathcal{A}$-sites only,
	\begin{equation}
		\mu_{i}=\begin{cases}
			\mu & i \in \mathcal{A} \\
			0 & i \in \mathcal{B}
		\end{cases},
	\end{equation}
	which might be experimentally more feasible, since it only requires a single normal metal layer. In what follows, we assume for simplicity that the Gilbert damping enhancement $\alpha_{\mathrm{sp}}$ is still present on all sites---such that $\alpha=\alpha_0+\alpha_{\mathrm{sp}}$ is constant throughout the system---but this is not a necessary requirement. After linearization and Fourier transforming, we obtain the same equation of motion \eqref{eq:nh:eom}, but with $\mathrm{i}\gamma\sigma_z\rightarrow \mathrm{i}\gamma(\sigma_0 + \sigma_z)/2$, such that there is only an imaginary mass on the $\mathcal{A}$-sites. 
	The long-wavelength excitations are then, up to first order in the dissipative terms, $\omega_{\bm k=0}=H-\mathrm{i}(\alpha H -\gamma/2)$ and thus the system is only stable if $\gamma/2<\alpha H$. There are therefore now two stability requirements: $\alpha_{\mathrm{sp}}\mu/2<\alpha H$ and $\mu < H$.

\begin{figure*}
		\centering
		\tikzsetnextfilename{asym}
		\includegraphics{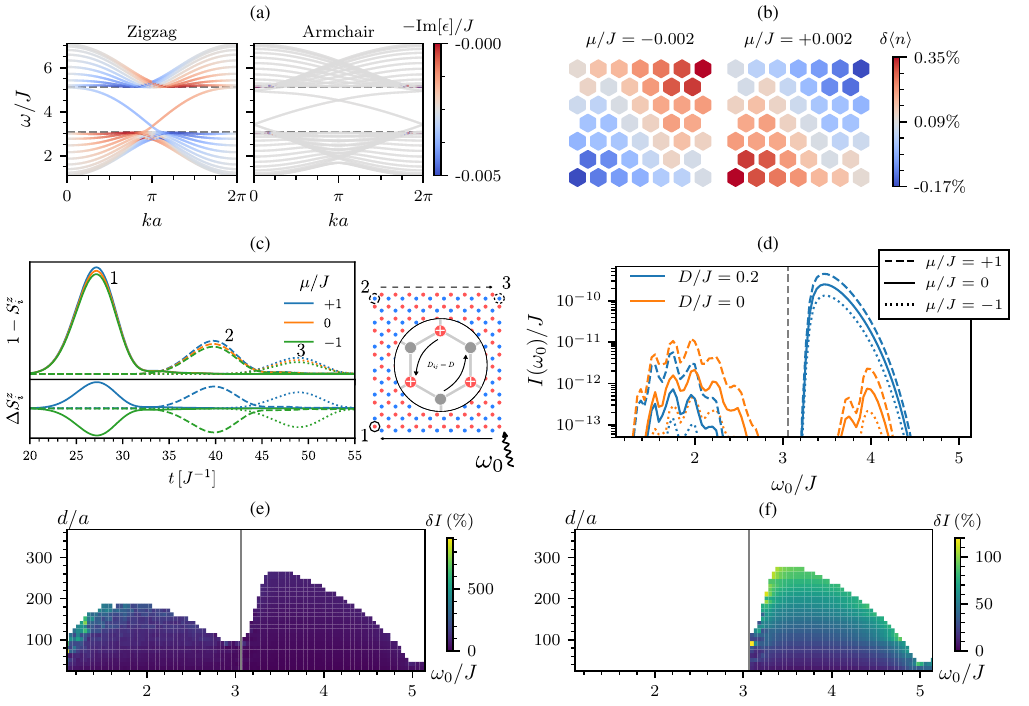}
		\caption{For the asymmetric spin-orbit torque only applied on the $\mathcal A$-sites, showing that \emph{asymmetric} spin-orbit torque exhibits the same features as the \emph{antisymmetric} spin-orbit torque. (a) The bandstructure [equivalent of \cref{fig:nh:periodic}]; (b) magnon density [equivalent of \cref{fig:hexagons-and-averaging}]; (c) numerical LLG simulations [equivalent of \cref{fig:nh:square}]; (d) transported spin current [equivalent of \subfigref{fig:magnon-transport}{(a)}]; (e) the relative amplification and (f) the relative amplification, only showing the edge mode [equivalents of \subfigref{fig:amplification-factor}{(a)}]. In (e), the colorscale is dominated by the bulk modes at low frequency. We therefore also show in (f) the relative amplification factor for only the edge modes, in order to highlight the similarity with the antisymmetric spin-orbit torque presented in the main text. \label{fig:asym} }
	\end{figure*}

	We first calculate the bandstructure for a zigzag and armchair nanoribbon, shown in \subfigref{fig:asym}{(a)}. Here we observe the same amplification of the right-moving edge modes compared to the left-moving edge modes [cf. Fig.~\ref{fig:nh:periodic} in the main text] for the zigzag nanoribbon. However, the damping correction is negative everywhere, due to the fact that $\mu_i>0$ everywhere---as opposed to the antisymmetric setup, where $\mu_i$ has opposite signs on different sublattices. We also calculate the magnon density, as shown in \subfigref{fig:asym}{(b)}, which displays the same features as the antisymmetric setup [cf. Fig.~\ref{fig:nh:heatmap} in the main text].
	
	We next reproduce the numerical LLG simulations in \subfigref{fig:asym}{(c)}. We observe the same amplification [cf. Fig.~\ref{fig:nh:square} in the main text], splitting the signal depending on the sign of the spin-orbit torque. However, because of the asymmetric nature of the spin-orbit torque, the signal is not canceled at the third detector, and is only slightly reduced. 
	

	Finally, we  show the amplification of the spin current for the asymmetric setup in \subfigref{fig:asym}{(d-f)}. We can clearly observe the same amplification of the topological edge states [cf. \cref{fig:magnon-transport,fig:amplification-factor} in the main text]. We thus conclude that the same robust amplification of the chiral edge modes can be achieved with an asymmetric spin-orbit torque. However, we also observe that we obtain some spurious amplification of the bulk modes at low frequencies, which is not necessarily suppressed by the finite disorder. These bulk modes dominate the colorscale in \subfigref{fig:asym}{(e)}, and we therefore also show in \subfigref{fig:asym}{(f)} the relative amplification only for frequencies corresponding to the excitation of edge modes. From \subfigref{fig:asym}{(f)} we conclude that we obtain the same disorder-protected spin transport as in the presence of an antisymmetric spin-orbit torque.
	
\end{document}